\documentclass[prl,superscriptaddress,twocolumn,preprintnumbers]{revtex4}

\usepackage{amsfonts}
\usepackage{amsmath}
\usepackage{amssymb}
\usepackage{bm}
\usepackage{dcolumn}
\usepackage{epsfig}
\usepackage{graphicx}
\usepackage{graphics}
\usepackage[latin1]{inputenc}
\usepackage{latexsym}
\usepackage{rotating}
\usepackage{hyperref}

\newcommand\be{\begin{equation}}
\newcommand\ba{\begin{eqnarray}}
\newcommand\ee{\end{equation}}
\newcommand\ea{\end{eqnarray}}

\begin{document}
\title{Towards a Loop Quantum Gravity and Yang-Mills Unification}

\author{Stephon Alexander}
\affiliation{Department of Physics and Astronomy, Haverford College, Haverford, PA 19041, USA}
\affiliation{Department of Physics, Princeton University, New Jersey 08544, USA}
\affiliation{Institute for Gravitation and the Cosmos, Department of Physics, Penn State, Universtiy Park, PA 16802, USA}

\author{Antonino Marcian\`o} 
\affiliation{Department of Physics and Astronomy, Haverford College, Haverford, PA 19041, USA}
\affiliation{Department of Physics, Princeton University, New Jersey 08544, USA}

\author{Ruggero Altair Tacchi}
\affiliation{Department of Physics, University of California, Davis, CA 95616, USA}

\preprint{}

\begin{abstract}

\noindent
We propose a new method of unifying gravity and the Standard Model by introducing a spin-foam model. We realize a unification between an $SU(2)$ Yang-Mills interaction and 3D general relativity by considering a $Spin(4)\sim SO(4)$ Plebanski action. The theory is quantized {\it \`a la} spin-foam by implementing the analogue of the simplicial constraints for the broken phase of the $Spin(4)$ symmetry. A natural 4D extension of the theory is shown. We also present a way to recover 2-point correlation functions between the connections as a first way to implement scattering amplitudes between particle states, aiming to connect Loop Quantum Gravity to new physical predictions.   
   
\end{abstract}

\maketitle

{\emph{Introduction}}.  One of the main challenges of high energy physics over the last few decades has been to provide a viable quantum theory of gravity that makes contact with experiment.  In this letter, following the perspective discussed in \cite{tutti}, we propose a theory that includes quantum gravity and Yang-Mills (YM) interactions as subgroups of an overall gauge unified theory.  Our approach relies on the non-perturbative quantization \emph{\`a la} Loop Quantum Gravity (LQG) of the theory in its initial phase. Then the theory is broken down, through an explicit symmetry breaking, to the general relativity (GR) and the YM parts. 

The theory is a spin-foam model, where the fundamental degrees of freedom are spin-networks and are endowed with quantum number representations of the entire gauge group. The spin-foam is defined as living in a $\mathcal{N}$\emph{D} manifold and the spin-network in its foliation, as usual in LQG. A method to compute the expectation value of Wilson loops of the YM and the GR fields is proposed. This is equivalent to the  \emph{n-point} function defined in \cite{MoRo} and the method relies on the boundary formalism \cite{Rovellibook, OeCoRo}.

So as to provide the underlying structure of our approach and avoid mathematical complexities, we will show  a non trivial Euclidean $\mathcal{N}$\emph{=3} case.  Remarkably, this simplified case provides an exactly soluble toy model which shows the emergence of a quantum theory of GR and YM interactions from the spin-foam quantization of the overall theory.  We establish exactly how the simplicity constraints, which in \emph{4D} are realized from Thiemann's procedure of the master constraint \cite{EPRL}, are connected to the emergence of the YM kinetic term. We then provide the reader with the holonomy representation \cite{BPM2} of the boundary propagator $W$ which encodes spin-foam dynamics, propose an extension of spin-network coherent states for both the GR and YM sectors and discuss the expectation value of the Wilson loops of the connections in the holomorphic representation \cite{BPM2}. 

{\emph{A spin-foam proposal towards unification}}. The theory is defined by implementing the following procedure: 

\emph{i)} the action $S$ is a modified Plebanski BF theory  that lives over a $\mathcal{N}$\emph{D} oriented smooth manifold; 

\emph{ii)} the action is invariant under a unified Lie group $G$, defining a principal $G$-bundle  $\mathcal{P}_{G}$;

\emph{iii)} the basic fields of the theory are a connection $A$ on $\mathcal{P}_{G}$, an $ad\!\!-\!\!\mathcal{P}_{G}$-valued $(\mathcal{N}-2)$-form $B$ on $\mathcal{M}_\mathcal{N}$ and a multiplet of scalar fields $\Phi$ on $\mathcal{M}_\mathcal{N}$;

\emph{iv)} we overcome the limitations of the Coleman-Mandula theorem for a curved spacetime, due to  an initial phase completely background independent, and only a following ``broken'' phase with an emergent metric, as explained in detail in \cite{Per} for a general class of models. In the broken phase all the standard implications of the theorem are recovered in the low energy limit; 

\emph{v)} we use the spin-foam implementation of the LQG dynamics \cite{Rovellibook}. The details of the spin-foam quantization are based on the discretization of the path integral for the $BF$ theory and on the consequent imposition on the quantized kinematical Hilbert space of the ``Plebanski-like'' constraints to the BF theory;

\emph{vi)} the generalized Hilbert space contains as {factors} the GR Hilbert space $\mathcal{H}^{\rm grav}_\Gamma$, the YM Hilbert space $\mathcal{H}^{\rm YM}_\Gamma$ and non trivial sectors related to the cosets generated by the symmetry breaking mechanism;

\emph{vii)} the asymptotic states expanded on spin-network basis elements do not necessarily  carry a simplicial interpretation \cite{Lew}. The spin-foam dynamics interpolates $1$-complexes, on which asymptotic states are supported \cite{LiSpi, BMMP}, and hence provides the proposal for a LQG predictive scattering process.

We believe that this proposal represents a robust and novel approach that implements LQG techniques in developing a unified theory. There are many peculiar subtleties in the $\mathcal{N}$\emph{=4} model, both conceptual and technical which may cloud fruitful progress. The issue, in fact, of dealing with a \emph{4D} spin-foam with $\{15 j\}_G$  re-coupling elements derived by the contraction of the intertwiners of the unification group $G$,  makes the explicit calculations particularly laborious. The presence of sectors associated to the GR or YM cosets,  { which will be pursued in future work, are very interesting but not necessary to show the first important elements of innovation of the proposal}. In addition, despite recent successes in the derivation of asymptotics for pure gravity in \emph{4D} \cite{Barrett}, { disagreement among experts on how LQG matter degrees of freedom should emerge has created some level of ambiguity as to the expectations for phenomenology.} 

Thus in this work we explore a simpler model that obviates, in a natural way, some of these difficulties. Nonetheless we are still able to show the richness of the enlarged spin-network Hilbert space and its proposed phenomenological interpretation.  In order to achieve this goal, we study a  Plebanski theory over a \emph{3D} oriented smooth manifold $\mathcal{M}_3$, over which we choose to consider a principal $Spin(4)$-bundle $\mathcal{P}_{Spin(4)}$. The basic fields of the theory are then a connection $A$ on $\mathcal{P}_{Spin(4)}$, an $ad(\mathcal{P}_{Spin(4)})$-valued $1$-form $B$ on $\mathcal{M}_3$ and a multiplet of scalar fields $\Phi_{ABC}$ on $\mathcal{M}_3$ {that is skew-symmetric in the indices}, with capital latin letters labeling indices in the adjoint representation of the algebra $\mathfrak{Spin}(4)\!=\! \mathfrak{su}(2) \!\times\! \mathfrak{su}(2)$.\\
The group $SO(4)\sim SU(2)\times SU(2)$ on a \emph{3D} manifold provides us with some evident simplifications:

\emph{i)} the two $SU(2)$ groups are naturally diagonal, making our model simpler than the full theory, but not trivial;

\emph{ii)} one $SU(2)$ will be interpreted as the GR sector, and is expected to be similar (at least as a limit) to the standard \emph{3D} LQG, a theory extensively studied; the other sector will be identified with an $SU(2)$ YM, which is the easiest non-abelian gauge theory we can write;

\emph{iii)} a $Spin(4)\sim SO(4)$  model is expected to share similarities with the standard \emph{4D} LQG (although the manifold dimensionality and the constraints are different);

{\emph{An explicit 3-dimensional model}}. We claim that both an $SU(2)$ YM  and GR can be unified in \emph{3D} by a modified $BF$ theory of the form
\begin{equation} \label{action}
S^{Pleb}\!=\!\frac{1}{G}\int_{\mathcal{M}_3} \!\!\!B^I \wedge F_I(A) - \Phi \cdot \mathcal{B} 
+ g\, \Phi \cdot \mathcal{B} \left( \Phi \cdot\Phi \right),\,
\end{equation} 
in which we have defined the  $3$-form $\mathcal{B}^{IJK}\!\equiv\! \!B^I\! \wedge B^J \!\wedge B^K$, denoted with $\cdot$ contraction of internal indices { and considered the sum over the internal index $I$ in the adjoint representation of $\mathfrak{spin}(4)$}. By variation of the action, manifestly Spin$(4)$ gauge invariant, Gau\ss\, law $\mathcal{D}_A\wedge B^I\!=\!0$ is recovered---$\mathcal{D}_A$ is the covariant derivative with respect to $A_I$. The ``field-strength constraint'' now reads $F_I=\Phi_{IJK} B^J \wedge B^K(1-g \Phi\cdot \Phi)$, while the generalization to the unified theory of those that are the simplicity constraints in the \emph{4D} $BF$-theory formulation of pure gravity 
\be \label{con}
\mathcal{B}^{IJK} \left( 1- g \, \Phi \cdot  \Phi \right) - 2 g \, \left( \Phi \cdot \mathcal{B}  \right) \Phi^{IJK}=0. 
\ee
The Spin$(4)$ symmetry of the theory is here broken by considering the {\it ansatz} on the decomposition of the multiplet of fields in $\Phi^{IJK}=\Phi^{ijk}\oplus \Phi^{abc}$, where the indices $ijk$ and $abc$ belong each one to a different $SU(2)\!\in\,$Spin$(4)$ subgroup, which is identified with the GR and YM theory, respectively.  We assume that the auxiliary field $\Phi^{ijk}$ is order $\sqrt{g\,}^{\,-1}$ and  $\Phi^{abc}$ is order $\sqrt{g\,}^{\,0}$, following the last Ref.~in \cite{tutti}. Expanding in $\sqrt{g}$ the equation of motion 
(\ref{con}), we easily find that the solution for the YM components of the multiplets are provided by $\Phi^{abc}=\lambda \epsilon^{abc}$ with $\lambda$ constant and of same dimension as $\sqrt{g}^{-1}$, and for the GR components by $\Phi^{ijk}=-(3\sqrt{2g})^{-1} \epsilon^{ijk}$. Pulling back the solution for $\Phi$ in the constraint (\ref{con}) provides (see \cite{AMT2}) the relation between the $\mathfrak{su}(2)$-valued components of $B^I$ 
\be \label{b}
B_{YM}=\gamma\, B_{GR}\,,
\ee  
that represents a second class constraint \cite{bookteitelboim} in the phase-space of the theory and in which $\gamma^{3}=3\lambda \sqrt{g/2}$. Equation (\ref{b}) implements the breaking of the Spin$(4)$ symmetry down to $SU(2)\!\times\!SU(2)$ in which the symmetry between the two subgroups is lost, and in this limit it gives the action for 3D gravity coupled to YM{, provided that (\ref{b}) is regarded as a constraint for the action defined by} 
\begin{eqnarray} \label{plin}
&\!\!\!\!\!\!\!\!S_{no\, \Phi}^{Pleb}[e, \omega, A, B]\!=\!\frac{1}{G}\!\int_{\mathcal{M}_3} \!\!\Big[e^i \wedge R_i(\omega) + B^a \wedge F_a(A) +\\
&\!\!\!\!\frac{+2\, \theta}{3 \sqrt{3g}} 
\sqrt{\!\epsilon^{\mu\nu\rho}\! \epsilon^{\alpha\beta\gamma}\! (e^i_{\alpha}e^i_{\mu}\!\!\!+\!\!B^a_{\alpha}B^a_{\mu}) (e^i_{\beta}e^i_{\nu}\!\!\!+\!\!B^a_{\beta}B^a_{\nu})(e^i_{\gamma}e^i_{\rho}\!\!\!+\!\!B^a_{\gamma}B^a_{\rho})}\, \Big].\nonumber
\end{eqnarray}
In (\ref{plin}) we have split the two subgroup components of the connection in $\omega^i$ (whose field strength is denoted as $R(\omega)$) for the GR sector and $A^c$ ($F(A)$ being the field strength) for the YM sector, and denoted the GR $\mathfrak{su}(2)$-valued $1$-form as $B^i_{GR}=e^i_\mu dx^\mu$, namely the triad, and the YM ones simply by $B^a$; the last term is equivalent to a cosmological constant term. The coupling constant $\theta$ is related to $g$ by $\theta(g)=\sqrt{1+\gamma^2}$. { Evaluating the action (\ref{plin}) in the $B^a$ field components of the stationary points (provided that these are subject to the constraint (\ref{b})), we recover 3D GR coupled to YM (see \cite{AMT2}):}
\be \nonumber
S_{no\, \Phi}^{\rm Pleb}=\frac{1}{G}\!\!\int_{\mathcal{M}_3} \!\!\!\!e^i \wedge R_i(\omega)  + \frac{3 \sqrt{\,3\,g}} {2\, \theta}  \int_{\mathcal{M}_3}  \!\!
\!\!F^a(A)\wedge \star F_a(A)\,.
\ee
Quantization {\it \`a la} spin-foam can be easily implemented in this context, following a standard recipe: 

\emph{i)}
the manifold is discretized by  introducing an oriented triangulation  $\Delta$ over $\mathcal{M}_3$, that is an abstract cellular complex constituted of points $p$, segments $s$ and triangles $t$. In the dual complex $\Delta^*$, constituted by  vertices $v$, edges $e$ and faces $f$, $n$-dimensional objects belonging to $\Delta$ are mapped in $(3-n)$-dimensional ones. 

\emph{ii)} It follows that each $SU(2)$ subgroup of the $B^{I}$ fields are smeared as algebra elements $B_s \equiv l_P^{-1} B^i_\mu l_s^\mu  \tau_i \sim   l_P^{-1} \tau_i \int_s B^i_\mu(\tilde{x}) dx^\mu$, $\tilde{x}\!\in\! s$ denoting a weighted point (with respect to the averaging procedure) along the segment $s$, $l_P$ the Planck length, $l_s^\mu$ an oriented averaged vector whose length is that of $s$ and $ \sigma_k=2i\,\tau_k$ Pauli matrices. 

\emph{iii)} Connection $A$ are smeared on the dual complex by associating to the discretization procedure group variables representing holonomies over edges $e\! \in\! \Delta^*$, namely $U_e \equiv e^{A_\mu^{ij} l_e^{\mu}}\sim e^{\int_e A }.$ These are conjugated variables to $B_s$ obeying canonical Poisson brackets. 

\emph{iv)} Loop quantization of the $SU(2)$-cotangent space over the spatial hypersurfaces of $\mathcal{M}_3$ proceeds constructing the Hilbert space of cylindrical functionals $\mathcal{H}_{\rm Cyl}$ \cite{PerezIntro}, over which holonomies are represented in a multiplicative way and fluxes are represented as left invariant derivative operators with respect to the connections \cite{Rovellibook}. 

\emph{v)} In $3D$ a basis is given by the eigenstates \cite{FER} of the area (volume in $4D$) and the length (area in $4D$) operators, {\it i.e.} the spin-network state basis $\psi_{\Gamma, j, \iota}$. Elements of this basis are supported on a graph $\Gamma\in\Delta^*$ and are labelled by spin $j$ of the irreducible representations (irreps) of each $SU(2)$ subgroup and by the intertwiner quantum number $\iota$. By construction, the elements $\psi_{\Gamma, j, \iota}$ are $SU(2)$ gauge invariant. Invariance under diffeomorphisms is implemented by considering topologically equivalent classes of graph $\Gamma$ over which $\psi_{\Gamma, j, \iota}$ are supported. The physical Hilbert space $\mathcal{H}_{\rm Phys}$ of the theory, implementing gauge and diffeomorphisms invariance \cite{LOST}, is then easily achieved by considering closure of $\mathcal{H}_{\rm Cyl}$ under the Ashtekar-Lewandowski (A-L) measure \cite{AL}. 

\emph{vi)} Realization of time re-parametrization encoded in the \emph{field strength} constraint (\emph{scalar constraint} for pure gravity in $4D$)
is implemented in a spin-foam setting by considering the discretization of the path integral of the theory \cite{Ro}.  An amplitude between the boundary graph $\Gamma$ of a $2$-complex (over which spin-foam is supported) yields the evolution of states over $\Gamma$. Consisting of two topological-$BF$-theories and $SU(2)$-symmetric sectors constrained by the additional symmetry breaking (\ref{b}), the theory results in a constrained sum over the two $SU(2)$ subgroups irreps, whose relation, derived by (\ref{b}), reads $j_{YM}\!=\!\gamma j_{GR}$. Denoting hence the SU$(2)$ subgroups irreps as $j_{GR}\!=\!j$ and $j_{YM}\!=\!\gamma j$, the partition function of the theory (\ref{action})  
\be  \label{kjk}
\mathcal{Z}^{\rm Pleb}_\Delta \!=\!\! \sum_{j_s, \,\gamma j_s} \prod_s {\rm dim}\,{j_s} \,\,{\rm dim}\,{(\gamma j)_s} \!\prod_\tau \{6\,j\}\,  \prod_{\tau'} \{6\,\gamma j\},
\ee
in which ${\rm dim}\,j$ stands for the dimension of the $j$ $SU(2)$ irreps 
and $ \{6\,j\}\,$ denotes the $6$-j symbol of $SU(2)$ recoupling theory. Notice that switching off $g$, and hence $\gamma$, accounts to obtain the sum from the Ponzano-Regge model, namely for $SU(2)$ topological $BF$ theory.  

\emph{Boundary propagator for one-vertex amplitude.} From (\ref{kjk}) we can extract the vertex amplitude and reformulate it in the holonomy representation \cite{BPM2}.  As a result, the vertex amplitude is achieved by performing an integration at each node over the gauge-group-elements $\tilde{G}\in$ \emph{Spin$(4)$}. If we are considering a one-vertex-amplitude, the integration over the bulk group element $G^{\rm bulk}$ of the two-complex is not necessary, as each $G^{\rm bulk}$ already represent \emph{Spin$(4)$} holonomies associated to the link $l$ of the boundary graph $\Gamma_4$. Then, assigning to any link $l$ a group-element $G_l\in$ \emph{Spin$(4)$}, 
\be \label{stivali}
\!\!W_v^{\rm Pleb}(G_l)\!=\! \int_{\rm Spin(4)^4} \prod \limits_{n=1}^{4} d\tilde{G}_n  \prod \limits_l \mathcal{K}_0 \left(\tilde{G}_{n_l} \, G_l \,\tilde{G}_{n'_l}^{-1}   \right),
\ee
where $\mathcal{K}_0=\mathcal{K}_t|_{t=0}$ and $\mathcal{K}_t$ denotes the propagation heat-kernel, whose heat-time is $t$ and that is expressed as a sum over the irreps of each \emph{SU$(2)$} subgroup of \emph{Spin$(4)$}: 
\be \nonumber \label{gatto}
\mathcal{K}_t(G)\!=\!\!\sum\limits_{j,\, \gamma j} {\rm dim} j\,{\rm dim} (\gamma j) \,e^{-j(j+1) \frac{t}{2}} {\rm Tr} \!\left[ \!\Pi^{(j, \gamma j)}\!(\tilde{G}_{n}G\tilde{G}^{-1}_{n'}) \!\right]\!.
\ee
The vertex amplitude (\ref{stivali}) provides the restriction of the boundary propagator to the tetrahedral graph $\Gamma_4\in\Delta^*$. This restriction can be thought to originate (see {\it e. g.} \cite{Carlo}) from the perturbative expansion in the coupling constant $\lambda$ of an appropriate Group Field Theory \cite{Frei} for the unified Plebanski theory here studied.

{\emph{Coherent spin-network states for the broken theory}.} Spin-network states for the broken phase of the full theory can be constructed generalizing \cite{LiSpi, BMMP} and references therein. Instead of considering only one SL$(2,\mathbb{C})$ group element for labeling coherent states (such as \cite{BMMP}), we must consider an element  $\mathbb{H}=H \times H' $ of ${\rm SL}(2,\mathbb{C})\otimes {\rm SL}(2,\mathbb{C})$. We assume that the two group elements $H$ and $H'$ carry the same information about the normals to the $1$-cells of the triangulation, {\it i.e.} to the segments $s$ bounding triangles. The ${\rm SL}(2,\mathbb{C})$ elements $H_l$ decompose as a complexification of \emph{SU$(2)$} elements by $H_l=n_{s(l)} e^{-iz_l \frac{\sigma_3}{2}}n_{t(l)}^{-1}$. In \emph{3D} each $H_l$ is hence labelled by two normals to the segment $s$, namely $n_{s(l)}$ and $n_{t(l)}$, whose relative rotation is achieved by a \emph{U$(1)$} subgroup of \emph{SU$(2)$}. For the element $H$ labeling the GR subgroup of the coherent states, the complex parameter $z_l=\xi+i\eta$ has the same meaning as in \emph{4D}: $\xi$ expresses the dihedral angle of a semiclassical Regge geometry, while $\eta$ the length of the $1$-simplices, {\it i.e.} the segments $s$. The $H'_l$ element labeling the YM subgroup can be thought as the necessary quantities to define a YM copy of the Regge geometry (as a YM lattice \cite{AMT2}). 

The complex parameters $z'_l$ of $H_l'$ are associated with the length of the YM lattice spacing, and is related to the flux though $s$ of the electric field $B_{YM}$. As a consequence of (\ref{b}), the flux of $B_{YM}$ is the $\gamma$ rescaling of the GR electric field flux. In a similar way, the GR dihedral angle is mapped, by multiplication by $\gamma^2$, in the equivalent dihedral angle of the YM lattice. This follows from (\ref{b}) and the expression of the extrinsic curvature in terms of $\xi$ (see {\it e.g.} \cite{RoSpe2}). As on the YM lattice  $\xi_\gamma=\xi\gamma^2$ represents the conjugated variable to the flux of the electric field, we can argue that $\xi_\gamma$ represents the index contraction of the gauge invariant field strength $F(A_{YM})$. Finally, coherent spin-network states read
\be \nonumber
\Psi_{\Gamma,\mathbb{H}_{l}}(G_{l})\!=\! \int_{\rm Spin(4)^4}\! \left( \prod \limits_{n} d\tilde{G}_n \right) \prod \limits_{l} \, \mathcal{K}_{t_{l}}\left(G_{l}, \tilde{G}_n \mathbb{H}_{l} \tilde{G}^{-1}_n \right)\,,
\ee
in which the heat kernel $\mathcal{K}_{t}$ has been specified above. 

\emph{Expectation value of product of holonomies.} The reconstruction theorem  \cite{Bar} ensures that gauge-invariant information about the principle fiber bundle $\mathcal{P}_{\rm Spin(4)}$ can be recovered from Wilson loops. Therefore the boundary formalism, developed in \cite{MoRo} and \cite{Carlo}, paves a way to compute the expectation value of the product of two holonomies, each belonging to a different \emph{SU$(2)$} subgroup of the theory. In the most straightforward setting, this expectation value will be calculated on the connected graph $\Gamma_4$, the tetrahedral spin-network. We evaluate Wilson loops $U_{\beta_{\bf x}}(h_l)$ and $U_{\beta'_{\bf y}}(h'_l)$, where $h_l$ and $h'_l$ are \emph{SU$(2)$} group-elements for each subgroup of \emph{Spin$(4)$}, and $\beta_{\bf x}$ and $\beta'_{\bf y}$ are loops with base points ${\bf x}$ and ${\bf y}$. For convenience, say that the two base points correspond to two nodes of $\Gamma_4$, and that the two loops bound two triangles sharing a segment. Within the Euclidean space $\mathcal{M}_3$ taken into account, we can think this graph to be embedded on the Regge submanifold that is the discretization of the boundary of a 3-ball, namely of $S^2$. The boundary propagator is described by $W_v(G_l)$, while the coherent states, representing the state over which the expectation value is computed, are given by $\Psi_{\Gamma,\mathbb{H}_{l}}(G_{l})$. Both of them are supported on $\Gamma_4$. At the first order in the GFT parameter $\lambda$ we can calculate
\be \label{a}
\mathcal{A}=\langle W_v (G_{l})| U_{\beta_{\bf x}}(h_l) U_{\beta'_{\bf y}}(h'_l)| \Psi_{\Gamma,\mathbb{H}_{l}}(G_{l})\rangle\,,
\ee
in which we use the inner product of the A-L measure \cite{AL} for each \emph{SU$(2)$} subgroup. This ensures gauge invariance and space-diffeoinvariance for (\ref{a}). The result is the sum over \emph{SU$(2)$} spin $j$ of the product of the expectation value of $U_{\beta_{\bf x}}(h_l)$ on the GR subgroup of $W_v (G_{l})$ and $\Psi_{\Gamma,\mathbb{H}_{l}}(G_{l})$, say $\tilde{\mathcal{A}} (j_i,s)$ its ``spin and intertwiner representation'', and of $U_{\beta'_{\bf y}}(h'_l)$ on the YM subgroup, say it $\tilde{\mathcal{A}}(\gamma j_i,s)$:
\begin{eqnarray} \label{tar}
&\!\!\!\mathcal{A} =\!\!\sum \limits_{j_l, \gamma j_l}\!\tilde{\mathcal{A}} (j_m,s_1) \!\prod \limits_{l} {\rm dim} j_{l} \,e^{-\frac{(j_{l}-j_{l}^0)^2}{2 \sigma^2_{l}}}  e^{-i \xi_{l} j_{l}}   \prod \limits_{n} \Phi_{\iota}(n_{l}) \times \nonumber\\
&\!\!\!\!\!\!\!\!\!\tilde{\mathcal{A}}(\gamma j_n,s_1)\!\prod \limits_{l'} {\rm dim} \gamma j_{l'} \,e^{-\frac{(\gamma j_{l'}- \gamma j_{l'}^0)^2}{2 \sigma^2_{l'}}} \!\!e^{-i \xi_\gamma^{l'} \gamma j_{l'}}  \! \prod \limits_{n} \Phi_{\iota_\gamma}(n_{l'}).
\end{eqnarray} 

In (\ref{tar}) $\iota$ ($\iota_\gamma$) denotes a trivalent intertwiner between irreps $j$ ($\gamma j$), the coefficients $\Phi_{\iota}(n_{l})$ and $\Phi_{\iota_\gamma}(n_{l})$ are the coherent intertwiner defined in \cite{LiSpi}, and finally ${\rm dim}\, j^0_{l}= \eta_{l}/t_{ab}$, ${\rm dim}\, \gamma j^0_{l}= \gamma \eta_{l}/t_{ab}$ and $\sigma^2_{l}=1/(2 t_{l})$. Each $\mathcal{A}$ is the the contraction of twelve Wigner $3 j$ symbols involving the six GR \emph{SU$(2)$} irreps $j$ (or YM $\gamma j$) labelling $\Gamma_4$ on the boundary of the interaction region. 
Eq. (\ref{a}) is the first step to implement the scattering of particle states in this research program, to connect LQG to physical predictions.
\begin{figure}[htb]
\begin{center}
\!\includegraphics[scale=0.2]{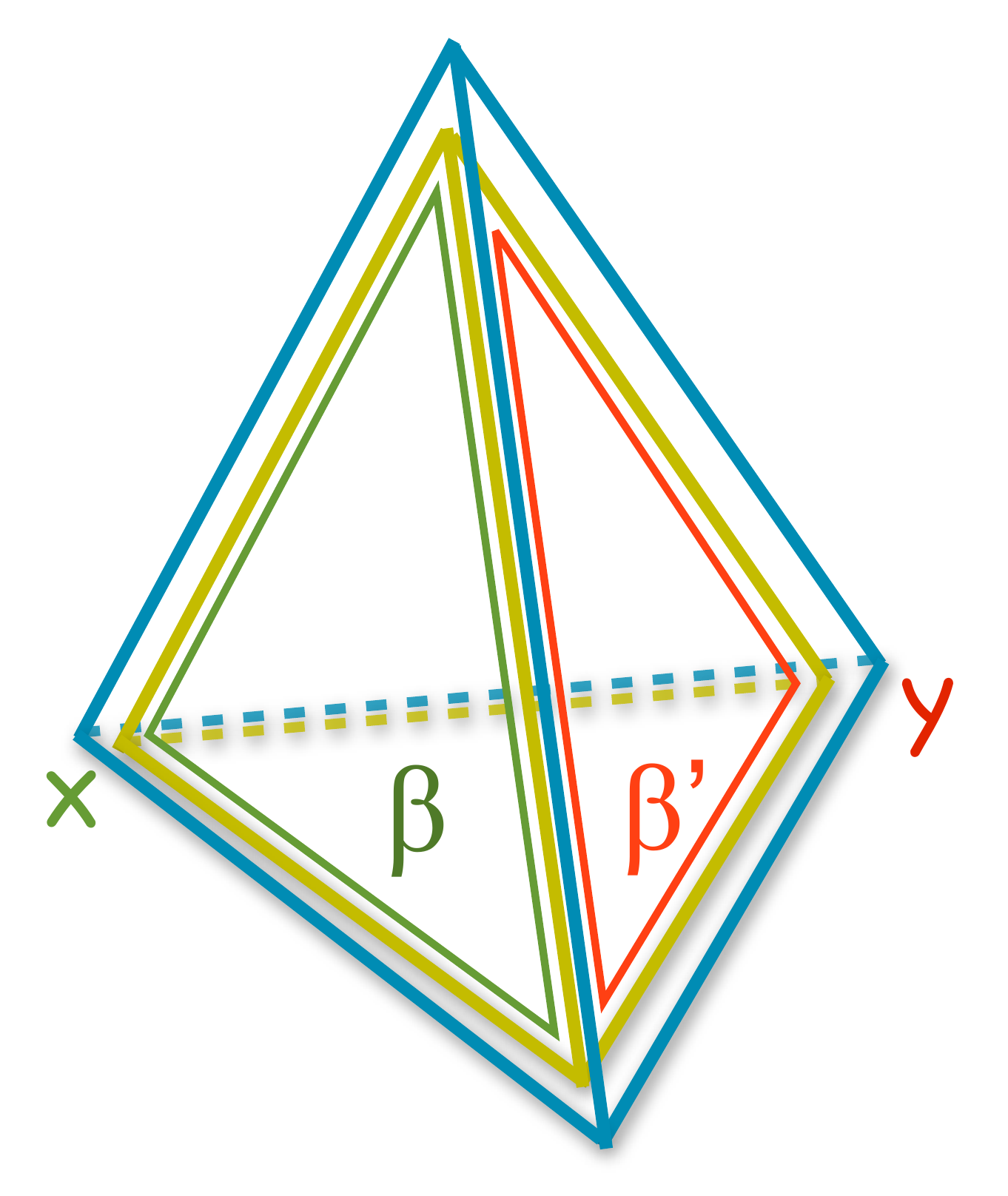} \hspace{5mm}\includegraphics[scale=0.28]{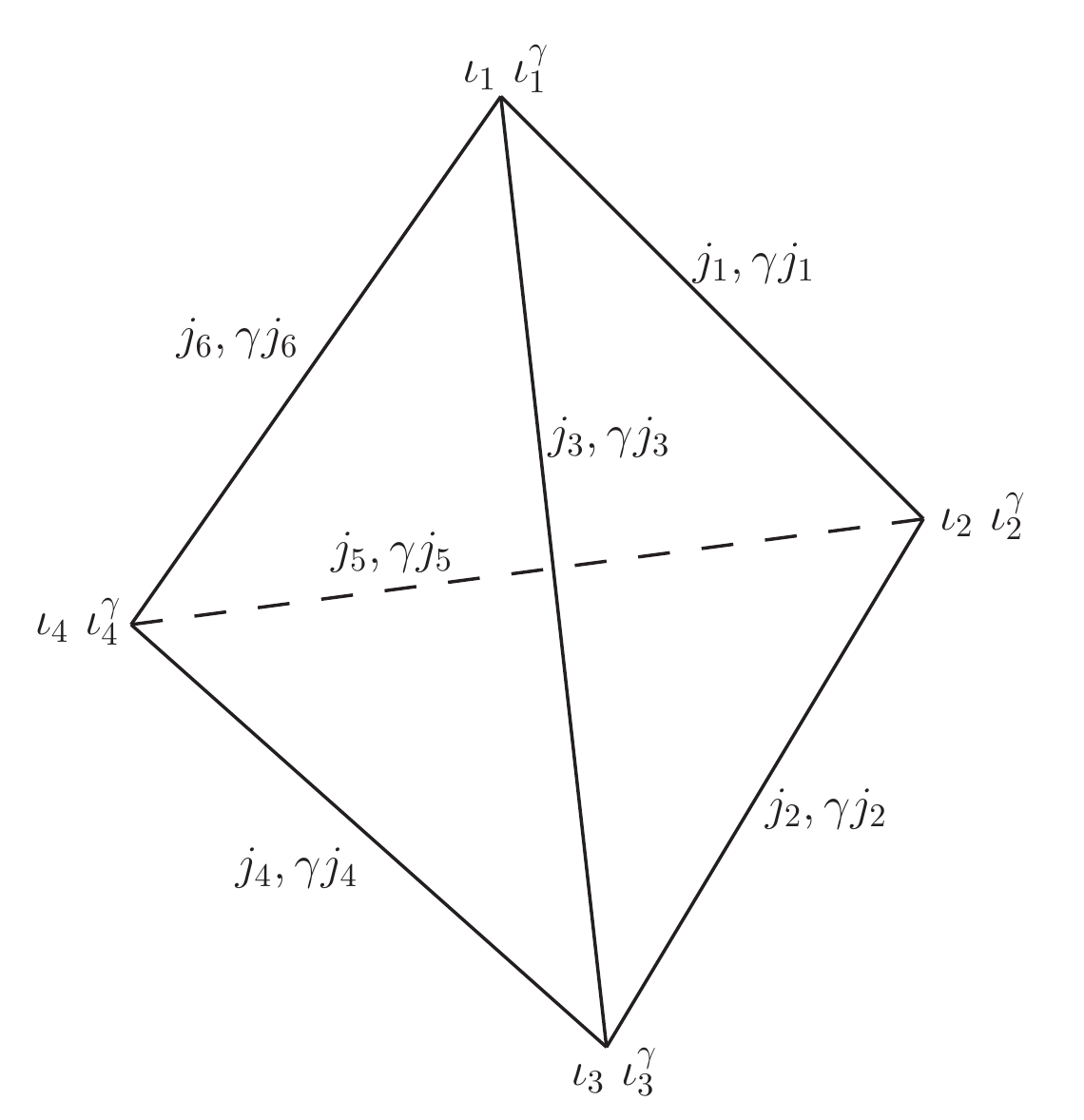}
\caption{Two $\Gamma_4$ and two loops $\beta_x$ and $\beta'_y$ on $S^2$, as in (\ref{a}) (left). $\Gamma_4$, colored with GR irreps $j$ and YM irreps $\gamma j$ (right).}
\end{center}
 \vspace{-4mm}
\end{figure}

 {\emph{Conclusions}}. We present a proposal for unifying gravity and Yang-Mills theory in LQG. The model offers exciting prospects for both theoretical and phenomenological development. Interesting work has been done in gravity and YM in \emph{3D} and in topological phases of matter with fractional statistics in \emph{3D} BF theory \cite{DJT-Choo}; it would be important to develop the model to compare it with these well known results. Although we believe that much can be understood in the dimensionally reduced case, the procedure is naturally implemented in \emph{4D} with no obvious obstacle if not for a more complex manipulability.

Finally, the proposed scattering amplitude provides large room for phenomenological predictions, especially after including (in future work) fermionic multiplets, and could be an important milestone for pushing LQG beyond its present limitations.
 
{\emph{Acknowledgements}}. We  thank G. Amelino-Camelia, C. Rovelli and L. Smolin for very stimulating discussions.

\end{document}